\documentclass[english]{article}
\usepackage[LGR,T1]{fontenc}
\usepackage[latin9]{inputenc}
\usepackage{color}
\usepackage{graphicx}
\usepackage{setspace}

\makeatletter

\DeclareRobustCommand{\greektext}{%
  \fontencoding{LGR}\selectfont\def\encodingdefault{LGR}}
\DeclareRobustCommand{\textgreek}[1]{\leavevmode{\greektext #1}}
\ProvideTextCommand{\~}{LGR}[1]{\char126#1}

\usepackage{babel}

\makeatother

\usepackage{babel}
\begin{document}
\title{\textcolor{black}{Testing Newton/GR, MoND and quantised inertia on
wide binaries}}
\author{\textcolor{black}{M.E.~McCulloch}\thanks{\textcolor{black}{University of Plymouth, PL4 8AA, mike.mcculloch@plymouth.ac.uk.}}\textcolor{black}{,
J.H.~Lucio}\thanks{\textcolor{black}{University of Plymouth, PL4 8AA.}}}
\maketitle
\begin{abstract}
\textcolor{black}{Wide binary stars are within the low-acceleration
regime in which galactic rotation curves deviate from Newtonian or
general relativistic predictions. It has recently been observed that
their rotation rates are similarly anomalous in a way that dark matter
cannot explain, since it must be smooth on these small scales to fit
galaxy rotation curves. Here, it is shown that Newtonian/GR models
cannot predict these wide binaries since dark matter cannot be applied.
It is also shown that MoND cannot predict these systems. However,
a model which assumes that inertia is due to Unruh radiation made
inhomogeneous in space by relativistic horizons (QI, quantised inertia)
can predict these wide binaries, and it has the advantage of not needing
an adjustable parameter.}
\end{abstract}

\section{\textcolor{black}{Introduction}}

\textcolor{black}{Zwicky (1933) first noticed that the motion of galaxy
clusters was too energetic to be held together by visible matter,
assuming Newtonian or general relativistic physics, and proposed the
existence of an invisible (dark) matter that provides the extra required
gravitational pull. A similar problem in disc galaxy rotation was
proven by the higher quality galaxy rotation curves obtained by Rubin
}\textit{\textcolor{black}{et al}}\textcolor{black}{.~(1980). Dark
matter is still the most popular explanation for the galaxy rotation
problem, but, after decades of searching, dark matter has not been
directly detected, though many efforts are ongoing, such as CDMS-II
(2009) and XENON10 (2009).}

\textcolor{black}{Milgrom (1983) proposed an alternative explanation
for galaxy rotation. He speculated that either 1) the force of gravity
increases or 2) the inertial mass ($M_{{\rm I}}$) decreases for the
low accelerations at a galaxy's edge. His empirical scheme, called
Modified Newtonian Dynamics (MoND), can fit galaxy rotation curves
and has the advantage of being less tunable than dark matter. However,
it does require tuning by one arbitrary parameter, the acceleration
$a_{0}$, it does not suggest a specific mechanism and it does not
predict the dynamics of galaxy clusters.}

\textcolor{black}{A theory has been proposed, see McCulloch (2007,
2013, 2016), in which inertia arises solely from a push on objects
by the quantum vacuum, which is made more intense by acceleration
(Unruh radiation) and made non-uniform in space by relativistic acceleration-dependent
Rindler horizons and able to push on matter. The theory predicts galaxy
rotation without dark matter and without any adjustment, see McCulloch
(2012, 2017a), and it implies that it is possible to produce new dynamics
by artificially creating horizons, damping the quantum vacuum, making
it inhomogeneous and able to push on objects, see McCulloch (2017b).
In QI the inertial mass becomes 
\begin{equation}
%M_{\rmI}=M_{\rmg}\left(1-\frac{2c^{2}}{|A|\Theta}\right)
M_{{\rm I}}=M_{{\rm g}}\left(1-\frac{2c^{2}}{A\Theta}\right)\label{QI_inertial_mass}
\end{equation}
where $m_{{\rm g}}$ is the gravitational mass, $c$ is the speed
of light, $A$ is the total acceleration of the object relative to
the fixed stars, and $\Theta$ is the distance to the co-moving cosmic
diameter, $8.8\times10^{26}$\,m. This is a generally-accepted estimate
of the cosmic diameter assuming that inflation has pushed objects
beyond the distance that we can now see (see Bars and Terning, 2009).
This represents the diameter as it is now and not when the light was
emitted from the horizon. For the derivation of Eq.\,(\ref{QI_inertial_mass})
see McCulloch (2007, 2016). QI successfully predicts galaxy rotation
without dark matter, see McCulloch (2012, 2017a), and the interesting
pattern noticed by Sanders and McGaugh (2002) that the anomalous behaviour
in galaxies begins at the radius where the acceleration of the stars
drops below the acceleration of $a_{0}\sim2\times10^{-10}\,{\rm m/s}^{2}$.
The problem is that galaxies do not provide a clean test since dark
matter can be `fitted' to also explain them.}

\textcolor{black}{The much simpler globular clusters were studied
by Scarpa }\textit{\textcolor{black}{et al.}}\textcolor{black}{\,(2007),
who observed the same change in behaviour at the critical acceleration.
Hernandez }\textit{\textcolor{black}{et al}}\textcolor{black}{.\,(2011),
and later with better GAIA DR2 (Data Release 2) data, Hernandez }\textit{\textcolor{black}{et
al}}\textcolor{black}{.\,(2018), provided a brilliantly simple crucial
experiment: they looked at the behaviour of wide-orbit binary stars
(for which the critical acceleration $a_{0}$ occurs at a separation
of about 7000\,AU or 0.03\,pc). Again, they found that the start
of anomalous behaviour occurs at the critical acceleration, not at
a distance, so it is difficult to explain the anomalies with dark
matter. Since dark matter cannot be applied to them at these small
scales, wide binary systems allow a purer comparison between competing
theories of motion and, as shown here, quantised inertia predicts
their behaviour better than MoND, without needing a tunable parameter
(unlike MoND). As a caveat, it should be noted that other studies
for example Banik (2019) claim that more wide binary data is needed
to make the results of Hernandez }\textit{\textcolor{black}{et al}}\textcolor{black}{.,
(2018) conclusive.}

\section{\textcolor{black}{Method: Newtonian/GR}}

\textcolor{black}{Consider two stars, each one of mass $M$, mutually
bound and orbiting. Applying Newton's second and gravity laws to one
of the stars, we get 
\begin{equation}
F=M_{{\rm I}}a=\frac{GM_{{\rm g}}M_{{\rm g}}}{d^{2}}
\end{equation}
where $G$ is the gravitational constant, $d$ is the separation between
the stars, $M_{{\rm I}}$ is the inertial mass of one star and $M_{{\rm g}}$
is its gravitational mass.}

\textcolor{black}{For a stable orbit gravity must be balanced by the
centrifugal, inertial force. Assuming simple circular motion, so that
$a=v^{2}/r=v^{2}/(d/2)=2v^{2}/d$, we get 
\begin{equation}
\frac{GM_{{\rm g}}M_{{\rm g}}}{d^{2}}=\frac{2M_{{\rm I}}v^{2}}{d}
\end{equation}
where $v$ is the orbital velocity of each star. In standard physics,
it is assumed that $M_{{\rm g}}=M_{{\rm I}}\,(\equiv M)$, and this
gives the Newtonian orbital velocity: 
\begin{equation}
v_{{\rm N}}=\sqrt{\frac{GM}{2d}}
\end{equation}
and finally we compute the relative velocity of the binary system
($\Delta v=2v$): 
\begin{equation}
\Delta v_{{\rm N}}=\sqrt{\frac{2GM}{d}}\label{eqn_DeltavN}
\end{equation}
}

\textcolor{black}{Since the speeds involved (around 400 m/s) are low
relative to the speed of light and gravitational forces are weak,
the predictions of general relativity (GR) are indistinguishable from
Newtonian dynamics for these cases.}

\section{\textcolor{black}{Method: MoND}}

\textcolor{black}{MoND assumes that for very low accelerations, either
the strength of gravity or inertial mass is modified, see Milgrom
(1983). Using inertial-MoND, Newton's second and gravity laws, replacing
the inertial mass using the simple MoND function, Gentile }\textit{\textcolor{black}{et
al}}\textcolor{black}{.\,(2011), and naming $M_{{\rm g}}$ as $M$,
we get 
\begin{equation}
F=M_{{\rm I}}a=\left(\frac{1}{1+\frac{a_{0}}{a+a_{{\rm g}}}}\right)Ma=\frac{GMM}{d^{2}}
\end{equation}
where we have added the external field effect assuming the acceleration
around the galaxy is, $a_{{\rm g}}\approx1\times10^{-10}\,\mathrm{m/s}^{2}$,
taking the lowest case from the Solar-system values determined by
Iorio, 2014 and Blanchet and Novak, 2010 (all the binaries used here
are within 100pc of the Sun, much smaller than the Sun's Galactocentric
radius of 8.2 kpc). So that, after some algebra: 
\begin{equation}
a^{2}+\left(a_{{\rm g}}-\frac{GM}{d^{2}}\right)a-\frac{GM(a_{{\rm g}}+a_{0})}{d^{2}}=0
\end{equation}
}

\textcolor{black}{% Assuming simple circular motion, so that $a = v^{2}/r = v^{2}/(d/2) = 2v^{2}/d$, we get
Applying $a=2v^{2}/d$, we get 
\begin{equation}
v^{4}+\left(a_{{\rm g}}d-\frac{GM}{d}\right)\frac{v^{2}}{2}-\frac{GM}{4}(a_{{\rm g}}+a_{0})=0
\end{equation}
}

\textcolor{black}{Using the quadratic equation (and keeping just real
and positive values) we get %\begin{equation}
%	v_{\rm MoND}=\sqrt{\frac{GM}{4d}\pm\frac{1}{2}\sqrt{\frac{G^{2}M^{2}}{4 d^2} + GMa_0}}
%\end{equation}
%and finally
\begin{equation}
%\Deltav_{\rmMoND}=\sqrt{\frac{GM}{d}\pm\sqrt{\frac{G^{2}M^{2}}{d^{2}}+4GMa_{0}}}
\Delta v_{{\rm MoND}}=\sqrt{\frac{GM}{d}-a_{{\rm g}}d+\sqrt{\left(a_{{\rm g}}d-\frac{GM}{d}\right)^{2}+4GM(a_{{\rm g}}+a_{0})}}\label{eqn_DeltavMoND}
\end{equation}
}

\section{\textcolor{black}{Method: Quantised Inertia}}

\textcolor{black}{Now, starting as before with Newton's second law
and gravity law, replacing the inertial mass using quantised inertia
(McCulloch (2007), see Eq.\,(\ref{QI_inertial_mass}), above) and
naming $M_{{\rm g}}$ as $M$ gives 
\begin{equation}
%F=M_{\rmI}a=M\left(1-\frac{2c^{2}}{|A|\Theta}\right)\frac{2v^{2}}{d}=\frac{GMM}{d^{2}}
F=M_{{\rm I}}a=M\left(1-\frac{2c^{2}}{A\Theta}\right)\frac{2v^{2}}{d}=\frac{GMM}{d^{2}}
\end{equation}
where $A$ is the modulus of the total acceleration of the orbiting
stars relative to the fxed stars, which includes $a=2v^{2}/d$\,
which is just the radial centrifugal component plus the non-centrifugal
acceleration of $2c^{2}/\Theta$. Therefore 
\begin{equation}
\frac{GM}{2d}=v^{2}-\frac{2c^{2}v^{2}}{\left(\frac{2v^{2}}{d}+\frac{2c^{2}}{\Theta}\right)\Theta}
\end{equation}
}

\textcolor{black}{Rearranging this, gives 
\begin{equation}
v^{4}-\frac{GM}{2d}v^{2}-\frac{GMc^{2}}{2\Theta}=0
\end{equation}
Using the quadratic formula (and keeping just real and positive values),
the relative velocity predicted by quantised inertia is 
\begin{equation}
%\Deltav_{\rmQI}=\sqrt{\frac{GM}{d}\pm\sqrt{\frac{G^{2}M^{2}}{d^{2}}+\frac{8GMc^{2}}{\Theta}}}
\Delta v_{{\rm QI}}=\sqrt{\frac{GM}{d}+\sqrt{\frac{G^{2}M^{2}}{d^{2}}+\frac{8GMc^{2}}{\Theta}}}\label{eqn_DeltavQI}
\end{equation}
}

\textcolor{black}{This formula has the advantage over the MoND formula
in that it includes $2c^{2}/\Theta$ instead of $a_{0}$ (the adjustable
MoND parameter), meaning that all the parameters in Eq.\,(\ref{eqn_DeltavQI})
are fixed and known, so it cannot be tuned to the data. It either
works or it does not, whereas Eq.\,(\ref{eqn_DeltavMoND}) MoND has
an arbitrary, tunable, parameter.}

\section{\textcolor{black}{Results}}

\textcolor{black}{The values input into these equations are as follows.
An average solar mass in the study of Hernandez }\textit{\textcolor{black}{et
al}}\textcolor{black}{.\,(2011) is $M=1\times10^{30}$\,kg, and
in Eq.\,(\ref{eqn_DeltavMoND}) the MoND fitting factor is set to
its original value $a_{0}=1.2\times10^{-10}\,\mathrm{m/s}^{2}$, % , and $a_0 = 2 \times 10^{-10}\mathrm{m/s}^{2}$. 
although it is true that some newer versions of MoND use a different
value ($a_{0}=2\times10^{-10}\,\mathrm{m/s}^{2}$), which gives an
expression practically equivalent to QI, Eq.\,(\ref{eqn_DeltavQI}).
The co-moving cosmic horizon used in Eq.\,(\ref{eqn_DeltavQI}) is
$\Theta=8.8\times10^{26}$\,m. For all the models we assumed an uncertainty
in mass %of $[_{-50\%}^{+100\%}$ 
ranging from $0.5M$ to $2M$, for MoND an uncertainty in $a_{0}$
ranging from $1.0\;\mathrm{to}\;2.0\times10^{-10}\,\mathrm{m/s}^{2}$
and for QI an uncertainty in $\Theta$ of $9\%$.}

\textcolor{black}{Fig.\,1 shows the separation in parsecs along the
$x$ axis from 0.007 to 4 pc and the relative orbital velocity of
the orbiting binaries along the $y$ axis in km/s. The crosses show
the rms relative velocity of the 83 wide binaries from the GAIA DR2
database as presented in Fig.\,5 of Hernandez }\textit{\textcolor{black}{et
al.}}\textcolor{black}{\,(2018). The grey area shows the uncertainty
in these results. The five bins from left to right contained 21, 24,
17, 8 and 13 pairs of stars respectively, so that the narrower vertical
error bars for the fourth bin is unexpected. As a check on our results
for Newton and MoND, they are roughly in line with those from the
detailed study of Banik and Zhao (2018) who found that at 20 kAU,
or 0.1 pc, the prediction of Newton was 0.15 km/s (compared to 0.2
km/s here) and for MoND it was 0.2 km/s (compared to 0.3 km/s here).
The difference in values could be due to the difference in assumed
mass. They assumed the total mass to be 1.5$M_{\odot}$, we assumed
2$M_{\odot}$.}

\textcolor{black}{The dotted lines show the expected Newtonian or
general relativistic velocity curve and its upper and lower uncertainty
bounds. Newton/GR significantly underpredicts the observed speed for
separations greater than about 0.3 pc (60,000 AU). So here general
relativity is falsified and dark matter cannot be used to save it.}

\textcolor{black}{The dashed lines show the prediction of MoND and
its upper and lower uncertainty bounds with the standard fitting parameter
of $a_{0}=1.2\times10^{-10}\,\mathrm{m/s}^{2}$. This model is very
similar to GR and also underpredicts the data for separations greater
than 0.3pc. This is due to the External Field Effect of MoND which
means that the binary stars still have a large acceleration because
of their orbit around the galactic centre, much higher than that needed
to show significant MoND effects.}

\textcolor{black}{The solid lines show the prediction of quantised
inertia and its upper and lower uncertainty bounds. QI alone agrees
with the data given the error bars. The advantage of QI here is that
it has no External Field Effect: the acceleration of the two stars
relative to the galaxy does not affect the inertial mass used in calculations
of their acceleration relative to each other. The other advantage
of quantised inertia is that it requires no tuning parameter, $a_{0}$.
It predicts this parameter itself. The data points all lie above the
prediction of QI, but still in agreement. It is possible that some
of the data has been contaminated by false binaries, though Hernandez
}\textit{\textcolor{black}{et al}}\textcolor{black}{. (2018) took
great care to avoid this. Nonetheless, Pittordis \& Sutherland (2019)
showed that false binaries are likely still an issue since a significant
fraction of systems have relative velocities too high to be plausible
in any gravity theory. These systems are likely to skew measures of
wide binary self-gravity based on rms relative velocities.}

\section{\textcolor{black}{Discussion}}

\textcolor{black}{The following is a more intuitive discussion of
the result. The binary stars' inertia is assumed in quantised inertia
to be caused by Unruh waves that are produced as the two stars accelerate
relative to other matter. This is due to the co-orbit of the two stars.
As more widely-separated binary stars are considered, their acceleration
relative to each other and to the rest of the matter in the universe
decreases, with the proviso that it remains above $2c^{2}/\Theta$.
Therefore the Unruh waves, that are assumed in quantised inertia to
determine their inertial mass, become longer, and a greater proportion
of them are disallowed by the cosmic horizon (a Hubble-scale Casimir
effect). This means that the inertial mass of the widely separated
stars decreases in a new way, and so they are able to orbit more quickly
than expected, without the centrifugal (inertial) forces separating
them (they remain bound).}

\textcolor{black}{As shown also by Hernandez }\textit{\textcolor{black}{et
al.}}\textcolor{black}{\,(2011, 2018), these data are in tension
with Newtonian or general relativity, since dark matter cannot be
added to these systems, as it must stay spread out at this scale to
fit galaxy rotations. MoND also does not predict these systems since
the External Field Effect makes it equivalent to GR. Only QI agrees
with the observed orbits within the uncertainty, and QI does it without
needing any tuning, which is a significant advantage.}

\section{\textcolor{black}{Conclusion}}

\textcolor{black}{Wide binary data from GAIA DR2 disagree with general
relativity since dark matter cannot be used in these cases.}

\textcolor{black}{MoND is also in disagreement with the data since
the External Field Effect makes its predictions similar to those of
general relativity.}

\textcolor{black}{Only quantised inertia (QI) agrees with the wide
binary data, and QI also has the advantage that it needs no fitting
parameter.}

\section*{\textcolor{black}{Acknowledgments}}

\textcolor{black}{Many thanks to the anonymous reviewers for their
helpful comments, and to DARPA grant HR001118C0125.}

\section*{\textcolor{black}{References}}

\begin{singlespace}
\textcolor{black}{CDMS Collaboration, (Ahmed Z.\,}\textit{\textcolor{black}{et
al}}\textcolor{black}{.), }\textit{\textcolor{black}{Phys. Rev. Lett.}}\textcolor{black}{,
102, 011301. }
\end{singlespace}

\textcolor{black}{Banik, I., H-S Zhao, 2018. Testing gravity with
wide binary stars like \textgreek{a} Centauri MNRAS, 480 (2), 2660
- 2688.}

\textcolor{black}{Banik, I., 2019. A new line on the wide binary test
of gravity. }\textit{\textcolor{black}{MNRAS}}\textcolor{black}{,
487, 5291 - 5303.}

\textcolor{black}{Bars, I., and J. Terning, 2009. Extra dimensions
in space and time. Springer.}

\textcolor{black}{Blanchet L., and J. Novak, 2010. External field
effect of modified Newtonian dynamics in the Solar system. MNRAS,
412(4)}

\textcolor{black}{Gentile, G., B.~Famaey and W.J.G.~de Blok, 2011.
THINGS about MoND. }\textit{\textcolor{black}{Astron.~Astrophys.}}\textcolor{black}{{}
527: A76.}

\textcolor{black}{Hernandez, X., M.A.~Jimenez, C.~Allen, 2011. Wide
binaries as a critical test of classical gravity. }\textit{\textcolor{black}{Astron.~Astrophys.}}\textcolor{black}{{}
arxiv:1105.1873v2}

\textcolor{black}{Hernandez, X., R.A.M.~Cortes, C.~Allen and R.~Scarpa,
2018. Challenging a Newtonian prediction through Gaia wide binaries.
}\textit{\textcolor{black}{MNRAS}}\textcolor{black}{.}

\textcolor{black}{Iorio, L., 2014. On the MOND External Field Effect
in the Solar System. Astrophysics and Space Science, 323(3):215-219}

\textcolor{black}{McCulloch, M.E., 2007. Modelling the Pioneer anomaly
as modified inertia. }\textit{\textcolor{black}{MNRAS}}\textcolor{black}{,
376, 338-342.}

\textcolor{black}{McCulloch, M.E., 2012. Testing quantised inertia
on galactic scales. }\textit{\textcolor{black}{ApSS}}\textcolor{black}{,
Vol. 342, No. 2, 575-578.}

\textcolor{black}{McCulloch, M.E., 2013. Inertia from an asymmetric
Casimir effect, }\textit{\textcolor{black}{EPL}}\textcolor{black}{,
101, 59001.}

\textcolor{black}{McCulloch, M.E., 2016. Quantised inertia from relativity
and the uncertainty principle. }\textit{\textcolor{black}{EPL}}\textcolor{black}{,
115, 69001.}

\textcolor{black}{McCulloch, M.E., 2017a. Galaxy rotations from quantised
inertia and visible matter only. }\textit{\textcolor{black}{Astrophys.~\&
Space Sci}}\textcolor{black}{., 362,149.}

\textcolor{black}{McCulloch, M.E., 2017b. Testing quantised inertia
on emdrives with dielectrics. }\textit{\textcolor{black}{EPL}}\textcolor{black}{,
118, 34003.}

\textcolor{black}{Milgrom M., 1983. A modification of the Newtonian
dynamics as a possible alternative to the hidden mass hypothesis.
$Astrophysical~Journal$, 270, 365. }

\textcolor{black}{Pittordis, C. and W. Sutherland, 2019. Testing modified
gravity with wide binaries in GAIA DR2. MNRAS (submitted).}

\textcolor{black}{Rubin, V.C., W.K.~Ford Jr, N.~Thonnard, 1980.
Rotational properties of 21 Sc Galaxies with a large range of luminosities
and radii, from NGC 4605 ($R=4$\,kpc) to UGC 2885 ($R=122$\,kpc).
}\textit{\textcolor{black}{Astrophysical Journal}}\textcolor{black}{,
238, 471.}

\textcolor{black}{Sanders R.H., S.S. McGaugh, 2002. }\textit{\textcolor{black}{ARA\&A}}\textcolor{black}{,
40, 263.}

\textcolor{black}{Scarpa, R., G.~Marconi, R.~Gimuzzi and G.~Carraro,
2007. }\textit{\textcolor{black}{A\&A}}\textcolor{black}{, 462, L9.}

\textcolor{black}{Zwicky,~F., 1933. Der Rotverschiebung von extragalaktischen
Nebeln. }\textit{\textcolor{black}{Helv.~Phys.~Acta}}\textcolor{black}{,
6, 110. }

\textcolor{black}{XENON10 Collaboration, }\textit{\textcolor{black}{Phys.~Rev.~D}}\textcolor{black}{,
80, 115005.}

\section*{\textcolor{black}{Figures}}

\textcolor{black}{\includegraphics[scale=0.4]{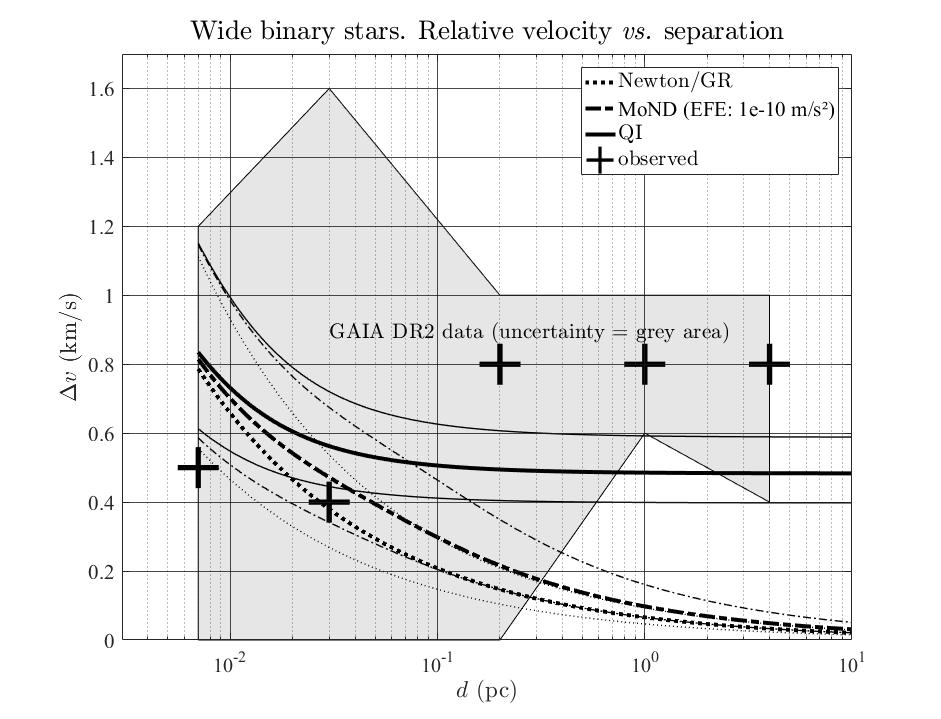}}

\textcolor{black}{Figure 1. The sky-projected relative velocities
of the wide binaries from Hernandez }\textit{\textcolor{black}{et
al}}\textcolor{black}{.\,(2018). The $x$-axis shows the separation
(in pc), and the $y$ axis shows the relative velocity (km/s). The
crosses and the grey areas show the observed velocities and possible
range. The three dotted curves show the predicted Newtonian or general
relativistic relative velocity and its upper and lower uncertainty
bounds. The three dot-dashed curves are the predictions from MoND
and its uncertainty bounds. The three solid curves represent the prediction
of quantised inertia (and its uncertainty bounds). QI is the only
model which agrees with the data.}
\end{document}